\documentclass[journal]{IEEEtran}
\usepackage{graphicx}  % Written by David Carlisle and Sebastian Rahtz
\begin{document}
\title{Single Junction and Intrinsic 
Josephson Junction Tunneling Spectroscopies 
of Bi$_{2}$Sr$_{2}$CaCu$_{2}$O$_{8+\delta}$}
\author{L. Ozyuzer, C. Kurter, J. F. Zasadzinski, K. E. Gray, D. G. Hinks and 
N. Miyakawa }% <-this % stops a space
\thanks{Manuscript received October 4, 2004.
        This research is supported by the U.S.\ Department of Energy, Basic 
Energy Sciences-Materials Sciences, under contract number 
W-31-109-ENG-38 and TUBITAK (Scientific and Technical Research 
Council of Turkey) project number TBAG-2031.  L.O. acknowledges support from 
Turkish Academy of Sciences, in the framework of the Young Scientist 
Award Program (LO/TUBA-GEBIP/2002-1-17).}% <-this % stops a space
\thanks{L. Ozyuzer is in the Department of Physics, Izmir Institute of Technology, 
Gulbahce Campus, Urla, 
TR-35430 Izmir, Turkey; phone: +90 232 7507518; fax: +90 232 7507509; 
e-mail: ozyuzer@iyte.edu.tr}
\thanks{C. Kurter is with the Izmir Institute of Technology, Izmir, 
Turkey}
\thanks{J. F. Zasadzinski is in the Physics Division, Illinois Institute of Technology, 
Illinois, USA}
\thanks{K. E. Gray and D. G. Hinks are in the Materials Science Division, Argonne 
National Laboratory, Illinois, USA}
\thanks{N. Miyakawa is with the Tokyo University of Science, Suwa, Japan}
% note the % following the last \IEEEmembership and also the first \thanks - 
% instance, "A\textbf{} \textbf{}B" will typeset as "A B" not "AB". If you want
% "AB" then you have to do: "A\textbf{}\textbf{}B"
\markboth{Journal of \LaTeX\ Class Files,~Vol.~1, No.~11,~November~2002}{Shell \MakeLowercase{\textit{et al.}}: Bare Demo of IEEEtran.cls for Journals}
% *** Note that you probably will NOT want to include the author's name in ***
% *** the headers of peer review papers.                                   ***

% If you want to put a publisher's ID mark on the page
% (can leave text blank if you just want to see how the
% text height on the first page will be reduced by IEEE)
%\pubid{0000--0000/00\$00.00~\copyright~2002 IEEE}

%\specialpapernotice{(Invited Paper)}

\maketitle

\begin{abstract}
Tunneling spectroscopy measurements are reported on optimally-doped and overdoped 
Bi$_{2}$Sr$_{2}$CaCu$_{2}$O$_{8+\delta}$ 
single crystals.  A novel point contact method is used to obtain 
superconductor-insulator-normal metal (SIN) and  
SIS break junctions as well as intrinsic Josephson junctions (IJJ) from nanoscale crystals. 
Three junction types are obtained on the same crystal to compare the quasiparticle peaks and higher 
bias dip/hump structures which have also been found in other surface probes such as scanning 
tunneling spectroscopy and angle-resolved photoemission spectroscopy.  However, our IJJ quasiparticle 
spectra consistently reveal very sharp conductance peaks and no higher bias dip structures.  
The IJJ conductance peak voltage divided by the number of junctions in the stack consistently leads 
to a significant underestimate of $\Delta$ when compared to the single junction values.  
The comparison of the 
three methods suggests that the markedly different characteristics of IJJ are a consequence of 
nonequilibrium effects and are not intrinsic quasiparticle features.
\end{abstract}

\begin{keywords}
High-T$_{c}$ superconductors, tunneling spectroscopy, intrinsic Josephson junctions.
\end{keywords}

% Will be ignored for other modes.
\IEEEpeerreviewmaketitle

\section{Introduction}
% \PARstart{A}{demo} file is ....
% \PARstart{A}{}demo file is ....
% \PARstart{T}{his demo} file is ....
\PARstart{W}{ith} conventional superconductors, it is relatively easy to fabricate stable  
superconductor-insulator-normal metal (SIN) and superconductor-insulator-superconductor  
(SIS) planar type junctions and the tunneling spectra from these have provided a direct  
measure of the temperature dependent energy gap, $\Delta$(T), and the electron-phonon 
interaction of 
strong-coupled BCS theory [1].  Yet for high temperature superconductors (HTS) no 
analogous planar junction technology exists.  Consequently, our understanding of the 
quasiparticle density of states (DOS) is obtained from mechanical junction methods.  
In Bi$_{2}$Sr$_{2}$CaCu$_{2}$O$_{8+\delta}$ (Bi2212) for example, SIN point contact 
tunneling (PCT) [2,3], SIS break junctions  
[2,3,4,5] and Scanning Tunneling Microscopy/Spectroscopy (STM/S) vacuum junctions [4,6,7] have
 all been achieved and the results are  
consistent.  The characteristic features of the Bi2212 junctions are quasiparticle peaks 
at the gap voltage that are both larger and broader than expected from a $d$-wave DOS as 
well as reproducible dip and hump structures at higher voltages which appear to be 
related to strong coupling effects [4,7,8].  The magnitude of the energy gap is 
consistently around 35-40 meV for optimally doped samples of Bi2212 and decreases 
(increases) with overdoping (underdoping) [3,5,8].  Another type of tunneling configuration 
has emerged which is unique to the layered, HTS cuprates, namely the intrinsic 
Josephson junctions (IJJ) between adjacent sets of Cu-O layers (e.g. the bi-layers of Bi2212) 
within the bulk crystal [9,10,11,12].  Various techniques have been developed so that stacks 
with only a few junctions in series can be studied.  The near perfect arrangement of such IJJ 
suggests that they will naturally lead to the intrinsic quasiparticle DOS of HTS, however, 
we will demonstrate here that the IJJ exhibit markedly different properties compared to the 
single-junction (SJ) methods and that the quasiparticle spectra of IJJ appear to be 
exhibiting nonequilibrium effects.

Interest has increased in IJJ because fabrication methods such 
as the patterning of small area mesas have led to a manageable number of such junctions 
in series  [10,11,12].  These are considered as c-axis tunnel junctions between sets of 
superconducting CuO$_{2}$ planes.   In the case of Bi2212, the CuO$_{2}$ bi-layers are only 0.3 nm  
thick and are separated by layers of Sr-O and Bi-O, which are 1.2 nm  thick and act as 
insulating or semiconducting spacers.     The measurements on small area mesas have 
generally been focused on the Josephson currents which can be studied at low bias 
voltages and the individual switching of each junction from the Josephson branch to the 
quasiparticle branch can be observed.   Surprisingly, this same pattern has also been 
observed at times using break junction methods which are intended to produce only a single 
SIS junction 
between pieces of the HTS crystal.  Presumably, in the case of Bi2212, nanoscale sized 
crystallites can result in the process of breaking the crystal and in some cases the I-V 
characteristics of these crystallites can be measured. 

The temperature and magnetic field evolution of current-voltage and conductance-voltage 
characteristics of IJJ have been intensively investigated. The study of the 
quasiparticle branch requires higher voltages and heating effects have been a problem.  
Various methods have been attempted to minimize such heating effects including  
intercalation and reduced junction area, although the latter still leads to the same 
power per unit area.  A short current pulse method has been used to minimize self-heating 
effects [12].  

Conventionally, the surface of bulk crystals have been patterned for 
different sizes and chemically etched or ion milled to obtain stack of junctions.  
Current and voltage leads have been connected using a passivation technique. This 
method allows a stable junction geometry so that temperature dependent measurements 
can be done very easily.  This is the most important advantage compared to PCT, break 
junctions and STM.  The disadvantage is that the chemical process steps may alter the 
intrinsic physical properties of the superconductor, especially the topmost Cu-O layer.  

  	In this paper, we will present a tunneling study of optimally doped and slightly  
overdoped Bi2212 as obtained from both SJ (SIN and SIS) and IJJ junction 
types.  All three 
junction geometries are obtained in-situ from a bare crystal at 4.2 K using a point 
contact method with a blunt Au tip.  We will compare results of the three distinct 
techniques and give an analysis of the quasiparticle tunneling spectra.  
It is argued that the strong dip/hump features found reproducibly in the 
SJ spectra are intrinsic and the absence of such features in the IJJ 
spectra are a consequence of heating or nonequilibrium processes 
such as quasiparticle injection.  Finally, the temperature 
dependence of SIS and IJJ tunneling 
conductances are compared and we will argue that this measurement 
also points toward the SIS junctions as revealing the intrinsic 
quasiparticle DOS.

\section{Experiment}
Single crystals of Bi2212 were grown by a floating zone technique.  As grown crystals 
are slightly overdoped and they are then annealed in Ar gas flow to obtain optimally 
doped T$_{c}$ values with an onset at 95 K.  Overdoping of samples has been achieved by 
annealing as grown crystals in O$_{2}$ gas flow.  The crystals of the present study have a 
T$_{c}$=82 K overdoped, and T$_{c}$=92 K near optimally doped with less than 2 K 
transition width 
found by ac susceptibility measurements.  Tunneling measurements were done using a 
continuous flow cryostat with a tunneling apparatus described in ref. [13].  Most of the 
tunneling data are taken at T=4.2 K.  Higher temperatures are obtained by adjusting the 
He gas flow and vaporizer temperature.  Samples are cleaved with a scotch  
tape to obtain clean surface of a-b plane and immediately mounted to the system where the 
counter electrode, Au tip, approaches along the c-axis of the crystal.    
	 
	 A novel method is used to form SIS break junctions and a detailed description 
can be found elsewhere [2].  First, SIN junctions are formed when the Au 
tip touches the non-conducting Bi-O/Sr-O surface bilayer. The  SIN
junctions provide an independent method for obtaining the superconducting 
gap.   These conductances exhibit quasiparticle peaks at 
eV$\sim\Delta$ and clearly show the dip and hump features above $\Delta$, 
most strongly observed in the occupied states 
of the DOS.  With increased 
pressure the tip breaks through the insulating surface layer forming an ohmic 
contact to the crystal and this contact forms a strong mechanical bond as 
well.  Further manipulation of the tip leads to a micro cleaving of Bi2212 
where a small crystallite is stuck to the Au tip and an SIS junction can 
be formed between this crystallite and the underlying crystal.   The convolution 
of the experimental SIN data with itself is then 
compared to the SIS conductances obtained from break junctions.  In all cases, the 
SIS conductances show quasiparticle peaks at 2$\Delta$ that are consistent with the $\Delta$ value 
obtained from the SIN junction.  This rules out the possibility of our SIS 
conductances being the result of multiple junctions in series.  Additionally, the SIS 
conductances also show a symmetric dip and weak hump feature as well as the presence of a 
Josephson current.  The magnitude of this current generally decreases as the junction 
resistance increases.  Sometimes breaking the sample with Au tip resulted multiple 
junctions which exhibited the typical IJJ characteristics found on 
fabricated Bi2212 mesas.  The I-V curve show multiple Josephson  
branches with hysteretic switching properties.

\begin{figure}
\centering
\includegraphics[bb=40 55 540 720, width=3in]{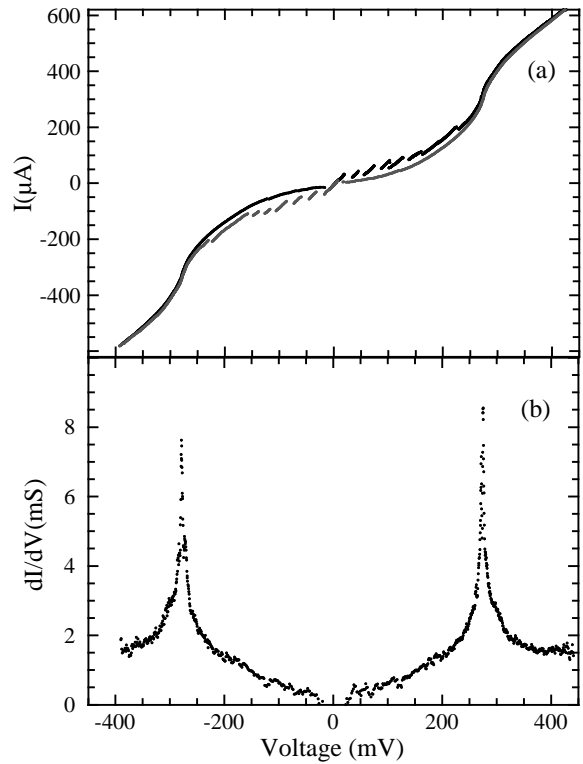}
% where an .eps filename suffix will be assumed under latex, 
% and a .pdf suffix will be assumed for pdflatex
\caption{(a) I-V and (b) dI/dV-V characteristics of IJJ made by 
point contact method on an overdoped 
Bi2212 crystal at 4.2 K. Josephson branches 
have been removed for clarity in dI/dV-V spectra.}
\label{fig_sim}
\end{figure}

\section{Results}
Figure 1(a) shows the I-V characteristics of an IJJ made by the point 
contact method on the overdoped Bi2212 crystal with 
T$_{c}$=82 K, on which SIN and SIS junctions were also obtained.  Multiple Josephson 
branches and hysteretic behavior are observed.   The gray curve represents bias 
sweeps from positive to negative voltage, and the black line is the reverse.  Since it is 
assumed that the 
number of branches in the I-V curve is the number of junctions, $n$, the dynamical 
conductance peak corresponds to $n$ times $\Delta_{IJJ}$.  We will call it  $\Delta_{IJJ}$ 
because of the 
substantially smaller value compared to $\Delta$ which is obtained from SIN and SIS junctions.  
In Fig. 1(a), 7 branches can be counted.  In Fig. 1(b), is shown the tunneling conductance 
of the same junction.  Here only the non-hysteretic branches of the I-V 
curve are shown for 
clarity.  The spectra show very narrow conductance peaks, V$_{p}$, at $\pm$274 mV without any features 
beyond $\pm n \Delta_{IJJ}$.  Below the conductance peaks, the tunneling conductance 
linearly increases 
up to $\pm$200 mV, which is similar to observation of ref. [14].  
  
\begin{figure}
\centering
\includegraphics[bb=55 55 540 480, width=2.9in]{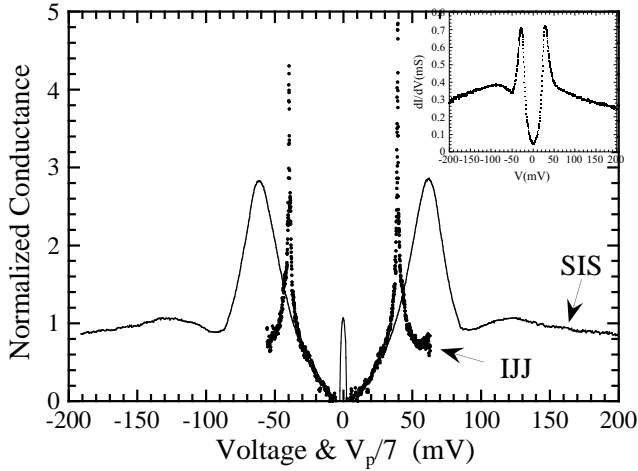}
% where an .eps filename suffix will be assumed under latex, 
% and a .pdf suffix will be assumed for pdflatex
\caption{Comparison of normalized conductance of SIS and IJJ on overdoped Bi2212 at 4.2 K. 
Inset shows tunneling conductance of a representative SIN spectra. 
All three spectra are obtained in a single run.}
\label{fig_sim}
\end{figure}

The inset of Fig. 2 shows SIN tunneling conductance characteristics 
obtained from the same crystal. The main panel of Fig. 2 shows the tunneling conductance 
of an SIS break junction 
at 4.2 K compared with the IJJ data.  The voltage 
bias axis of IJJ data is divided by 7 because of 7 branches, so both junction 
types can be shown on the same figure. 
The conductance peaks of SIS junction correspond to $\pm$2$\Delta$, 
$\pm$62 meV.  Furthermore, Josephson current peak at zero bias indicates 
the superconducting nature of the observed energy gap magnitude.  
An important difference between the two quasiparticle curves is 
the location and shape of the dynamical conductance peaks.  Convolution of 
the SIN data to generate an SIS conductance leads to a shape consistent with 
the measured SIS data.  IJJ displays 
sharp peaks at $\pm$39 mV which 
corresponds to 2$\Delta_{IJJ}$.  Thus the IJJ leads to an underestimate of $\Delta$ and $\Delta_{IJJ} \ne \Delta$.  
Excessive heat developed in the IJJ stack is the likely cause of the lower gap 
value.  This leads to an effective temperature of the stack which can be 
significantly larger than the bath temperature.  However, the ratio of 
$\Delta_{IJJ}$/$\Delta$ is around 0.65 which is close to the value of 
0.6 found in the 
theory of Owen and Scalapino [15] that considers the nonequilibrium 
electron distribution caused by excess quasiparticle injection.  Thus such injection effects
might be playing a role in the IJJ 
quasiparticle characteristics.    
Since the barrier strength is weak between CuO$_{2}$ planes, 
overinjection of quasiparticles may cause a nonequilibrium state [16].  Interestingly, 
near zero bias both junctions show a similar shape indicative of an anisotropic 
gap.   The  IJJ junction characteristics 
deviate for larger bias, again suggesting the development of a 
nonequilibrium state.

Another striking difference shown in Fig. 2 involves the higher bias 
dip/hump features which are completely absent in all of the IJJ studied.  
Since the dip/hump features are reproducibly observed in the SIN and SIS 
junctions as well as in STM/S spectra, they must be considered to be 
intrinsic properties of the electron DOS in the superconducting state.  
Their absence in the IJJ spectra is consistent with the loss of 
superconductivity in the IJJ at high voltages due to a combination of 
heating and injection effects.  Recently, the local temperature of the IJJ stacks 
have been obtained using a thin film thermocouple which is evaporated top of 
the mesa [17].  This technique shows that the local temperature of the mesa can 
exceed T$_{c}$ for junction bias voltages near the gap voltage.  This is 
consistent with the absence of dip and hump spectral 
features except when short-pulse methods are applied [12] or Bi2212 is intercalated 
with HgBr$_{2}$ [11].

\begin{figure}
\centering
\includegraphics[bb=40 55 540 650, width=3in]{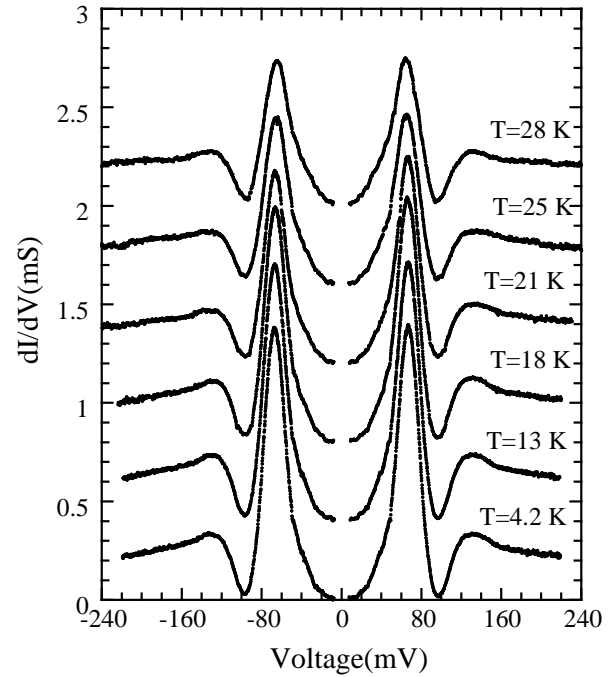}
% where an .eps filename suffix will be assumed under latex, 
% and a .pdf suffix will be assumed for pdflatex
\caption{Temperature dependece of tunneling conductance for a SIS junction. 
Each spectra is shifted 0.4 mS from each other and Josephson current peaks are deleted for clarity.}
\label{fig_sim}
\end{figure}

We now turn to the temperature evolution of the gap features. Figure 3 shows 
temperature evolution of tunneling spectra from an SIS break junction  
on optimal doped Bi2212 with T$_{c}$=92 K.  Josephson current peaks are 
deleted for clarity.  In conventional, 
$s$-wave superconductors, the temperature dependence of the energy gap  
magnitude, $\Delta$(T), is well understood, both theoretically and experimentally.   
$\Delta$(T) is nearly constant up to T$\sim$T$_{c}$/4.  In the vicinity of T$_{c}$, 
it closes to zero
where it is proportional to (1-T/T$_{c}$)$^{1/2}$.  High 
temperature superconductors exhibit properties which might affect this T 
dependence.  One such property is the 
presence of a pseudogap well above T$_{c}$, up to a characteristic temperature T*, that has 
been detected by a number of experimental techniques, such as 
in-plane resistivity, ARPES, specific heat, and 
NMR.  In addition to these experimental techniques, STM, break junction 
and planar tunnel junctions have also observed that the pseudogap exists as a 
depression in the DOS of Bi2212 above T$_{c}$.  Irrespective of the 
presence of a pseudogap or its origin, the superconducting gap should 
exhibit a weak T dependence far below T$_{c}$.

Because of the low zero bias conductance and sharp conductance peaks in Fig. 3, we can directly  
find energy gap from conductance peaks which correspond to $\pm$2$\Delta$.  For this junction,  
2$\Delta$=$\pm$67 meV.  The data also show very strong dip features (almost zero conductance) 
which  
is consistent with a previous break junction tunneling study [8] of similar crystals with 
T$_{c}$=95 K.  
Tunneling conductance peaks smear when the temperature is raised from 4.2 K to 28 K, however  
position of peaks hardly moves at all, consistent with a nearly constant 
superconducting gap as expected from conventional BCS theory.   This is illustrated 
in inset of Fig. 4 where the SIS gap (triangles) is compared with the BCS 
prediction (line).  
The Bi2212 gap drops slightly more rapidly than the BCS prediction but this 
can likely be accounted for by the d-wave gap symmetry in Bi2212.  Figure 4 
also shows temperature dependence of 
tunneling conductance of an IJJ obtained from the same Bi2212 crystal.    
The hysteretic branches are displayed because the bias sweep direction is a decreasing 
voltage.   The conductance peak voltage rapidly decreases with increasing temperature.   
In addition, the branches also 
move and that is seen in previous studies of IJJ [11].  The details of strong 
temperature dependencies of IJJ conductance peak can be seen in inset of Fig. 
4 (diamonds) along with data from 20x20 $\mu$m mesa (filled circles) fabricated on a similar 
Bi2212 crystal.  What is clearly evident is that IJJ spectra either from 
point contact generated stacks or patterned mesas show a consistent, rapid decrease of 
the conductance peak with temperature.  This is contrary to the SIS gap 
evolution (Fig. 3) and to the BCS prediction.  This is a further indication that the 
conductance peak in IJJ is not revealing the true quasiparticle gap.

In summary we have compared the quasiparticle spectra from single junction 
methods on Bi2212 with those of IJJ, including the temperature dependence.  
The characteristics of the IJJ spectra are very sharp conductance 
peaks, a reduced superconducting gap and the absence of dip/hump features.  
The conductance peak also exhibits an anomalously rapid decrease with 
temperature.  All of these features indicate the IJJ spectra are affected 
by non-equilibrium processes.

\begin{figure}
\centering
\includegraphics[bb=55 55 540 580, width=2.85in]{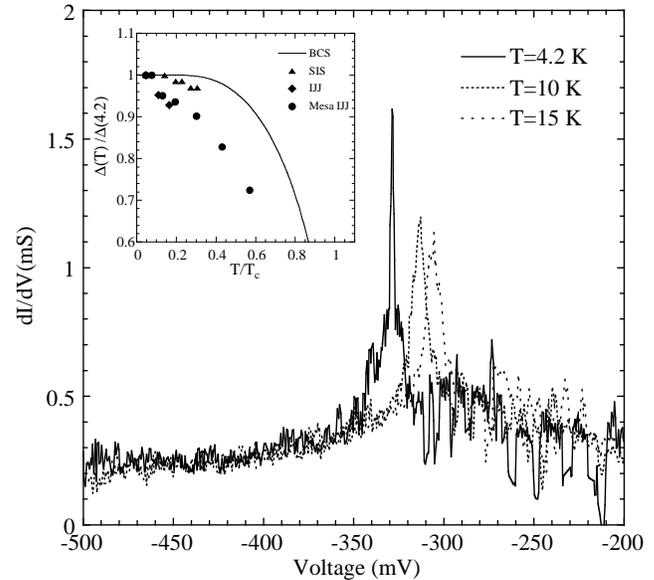}
% where an .eps filename suffix will be assumed under latex, 
% and a .pdf suffix will be assumed for pdflatex
\caption{Temperature dependence of tunneling conductance for IJJ.  The inset shows 
normalized energy gap versus normalized temperature for various junction types.}
\label{fig_sim}
\end{figure}


% http://www.ctan.org/tex-archive/macros/latex/contrib/supported/IEEEtran/bibtex...
%\bibliographystyle{IEEEtran.bst}

\end{document}